\documentclass[aps,noshowkeys]{revtex4}
\usepackage{epsfig}
\usepackage{graphicx}
\usepackage{bm}
\usepackage{amssymb,amsmath,euscript,esint,mathtools}
\begin{document}
\newcommand{\Br}[1]{(\ref{#1})}
\newcommand{\Eq}[1]{Eq.\ (\ref{#1})}
\newcommand{\frc}[2]{\raisebox{1pt}{$#1$}/\raisebox{-1pt}{$#2$}}
\newcommand{\frcc}[2]{\raisebox{0.3pt}{$#1$}/\raisebox{-0.3pt}{$#2$}}
\newcommand{\frccc}[2]{\raisebox{1pt}{$#1$}\big/\raisebox{-1pt}{$#2$}}
\title{$\mathcal{P}$,$\mathcal{T}$-odd Faraday rotation on atoms and molecules in intra-cavity absorption spectroscopy as an alternative way to search for the $\mathcal{P}$,$\mathcal{T}$-odd effects in nature}
\author{D. V. Chubukov$^{1,2}$, L. V. Skripnikov$^{1,2}$, L. Bougas$^{3}$, and L. N. Labzowsky$^{1,2}$}
\affiliation{$^1$ Department of Physics, St. Petersburg State University,
7/9 Universitetskaya Naberezhnaya, St. Petersburg 199034, Russia \\
$^2$Petersburg Nuclear Physics Institute named by B.P. Konstantinov of
National Research Centre ``Kurchatov Institut'', St. Petersburg, Gatchina 188300, Russia \\
$^3$Helmholtz Institut Mainz, Johannes Gutenberg-Universit\"{a}t, Staudingerweg 18, 55128 Mainz, Germany \\
}

\begin{abstract}
Present limit on the electron electric dipole moment ($e$EDM) is based on the electron spin precession measurement. We propose an alternative approach – observation of the $\mathcal{P}$,$\mathcal{T}$-odd Faraday effect in an external electric field on atoms and molecules using cavity-enhanced polarimetric scheme in combination with molecular (atomic) beam crossing the cavity. Our calculations of the effective electric fields and theoretical simulation of the proposed experiment on Tl and Pb atoms, PbF, YbF, ThO, and YbOH show that the present limit on the $e$EDM can be improved by 6-7 orders of magnitude.
\end{abstract}

\maketitle
\section{Introduction}
The existence of the electric dipole moment (EDM) for any particle or closed system of particles violates both the space parity ($\mathcal{P}$) and time-reversal ($\mathcal{T}$) symmetries~\cite{Khrip91,Gin04,Saf18}. Up to date the most stringent experimental limitations for the particles' EDMs are obtained for the electron ($e$EDM) due to its great enhancement in heavy atoms and diatomic molecules. The most restrictive $e$EDM bounds were established in experiments with a ThO molecule ($d_e<1.1\times 10^{-29}$ $e$ cm \cite{ACME18}). Previously, accurate results were obtained on the Tl atom \cite{Reg02}, YbF molecule~\cite{Hud11} and HfF$^+$ cation~\cite{Cair17}. For extraction of the $e$EDM values from the experimental data, accurate theoretical calculations are required. These calculations were performed for Tl~\cite{Liu92,Dzuba09,Por12,Chub18}, for YbF~\cite{Quiney:98, Parpia:98, Mosyagin:98, Abe:14}, for ThO~\cite{Skripnikov:13c,Skripnikov:15a,Skripnikov:16b,Fleig:16}, and for HfF$^+$~\cite{Petrov:07a,Skripnikov:17c, Fleig:17, Petrov:18b}. In the same experiments it is possible to search for another $\mathcal{P}$,$\mathcal{T}$-odd effect: $\mathcal{P}$,$\mathcal{T}$-odd electron-nucleus interaction~\cite{San75,Gor79,Koz95}. Effects from this interaction and from $e$EDM can be observed in an external electric field and cannot be distinguished in any particular atomic or molecular experiment. However, they can be distinguished in a series of experiments with different species \cite{Bon15,Skripnikov:17c}. The magnitude of the electron-nucleus $\mathcal{P}$,$\mathcal{T}$-odd interaction can be conveniently expressed via the equivalent $e$EDM $d_e^{\text{eqv}}$ which leads to the same linear Stark shift of atomic levels in the same external electric field.

Theoretical predictions of the $d_e$ value are rather uncertain. Within the Standard Model (SM) none of them  promises the $e$EDM value larger than $10^{-38}$ $e$ cm, however, predictions of the SM extensions are many orders of magnitude larger~\cite{Engel2013}. Various models for the $\mathcal{P}$,$\mathcal{T}$-odd interactions within the SM framework are discussed in Refs. \cite{Pos14,Chub16}.
In modern experiments for the $\mathcal{P}$,$\mathcal{T}$-odd effects observation in atomic and molecular systems either the shift of the magnetic resonance \cite{Reg02} or the electron spin precession \cite{Hud11,ACME18,Cair17} in an external electric field is studied. Due to a very large gap between the current experimental bound and the maximum SM theoretical prediction alternative methods for the observation of the $\mathcal{P}$,$\mathcal{T}$-odd effects are of interest. In Refs. \cite{Baran78,Sush78} it was mentioned the existence of the effect of optical rotation of linearly polarized light propagating through a medium in an external electric field \--- the $\mathcal{P}$,$\mathcal{T}$-odd Faraday effect. The possibility to observe it was studied theoretically and experimentally (see the short review \cite{Bud02}). Recently, a possible observation of the $\mathcal{P}$,$\mathcal{T}$-odd Faraday effect by the intracavity absorption spectroscopy (ICAS) methods \cite{Boug14,Baev99,Dur10} using atoms was considered \cite{Chub17}. In Ref. \cite{Boug14} an experiment on the observation of the $\mathcal{P}$-odd optical rotation in Xe, Hg, and I atoms was discussed. The techniques \cite{Boug14} is very close to what is necessary for the $\mathcal{P}$,$\mathcal{T}$-odd Faraday effect observation. In Refs. \cite{Chub18,Chub19:1,Chub19:2} an accurate evaluation of this effect oriented to the application of the techniques \cite{Boug14} were undertaken for the atomic case. In the present paper we consider both atomic and molecular systems for the beam-based ICAS $\mathcal{P}$,$\mathcal{T}$-odd Faraday effect observation (see below).

As it was shown in earlier works \cite{Khrip91}, heavy atoms and molecules containing such atoms are very promising systems to search for the $\mathcal{P}$,$\mathcal{T}$-odd effects. For the case of $\mathcal{P}$,$\mathcal{T}$-odd Faraday effect such systems should also satisfy the following requirements.

 i)
 The transition frequencies should belong to the optical (or near infrared) region. 
 
 ii)
 The natural linewidth of the chosen transitions (the collisional width is negligible for beam-based experiments) should be as small as possible otherwise the $\mathcal{P}$,$\mathcal{T}$-odd Faraday signal will be suppressed. For this reason the most suitable are the M1 transitions from the ground to metastable state, e.g.  $6p_{1/2} \rightarrow 6p_{3/2}$ in Tl and $6p^2 (1/2,1/2)_0 \rightarrow 6p^2(3/2,1/2)_1$ in Pb considered in Ref. \cite{Chub18}. 

 iii)
 The applied electric field $\mathcal{E}_{\text{ext}}$ should be close to the saturating field $\mathcal{E}_{\text{sat}}$, i.e. the field which begins to repulse the levels with opposite parity. For atoms $\mathcal{E}_{\text{sat}}$ can be never reached in the laboratory, therefore $\mathcal{E}_{\text{ext}}$ should be chosen as large as possible (the largest achievable field is $10^5$ V/cm \cite{Reg02}). However, such a field can be created only within the space of about 1 cm. Since the usual atomic beam diameter is of the order of several centimeters, the $\mathcal{P}$,$\mathcal{T}$-odd Faraday experiments with atoms look possible as a combination of ICAS and atomic beam experiments. 
For diatomic molecules with total electron angular momentum projection on molecular axis, $\Omega$, equal to $1/2$ such as PbF, YbF $\mathcal{E}_{\text{sat}}$ is about $10^4$ V/cm. Diatomic molecules with $\Omega > 1/2$ can be polarized at very small external electric fields due to closely lying levels of opposite parity (so-called $\Omega$-doubling). The importance of the use of $\Omega$-doubling for the search of $\mathcal{P}$- and $\mathcal{P}$,$\mathcal{T}$-odd effects was noted in Refs. \cite{Lab77,Sush78,Gor79}. Finally, linear three-atomic molecules with $\Omega = 1/2$ are very promising systems to search for the $\mathcal{P}$,$\mathcal{T}$-odd effects as they can be easily polarized due to the vibrational $l$-doubling effect~\cite{Koz17}.

We can imagine an ICAS-beam experiment for the $\mathcal{P}$,$\mathcal{T}$-odd Faraday rotation observation as follows. An atomic (molecular) beam crosses the cavity in transverse direction. Within a cavity it meets a laser beam travelling along. The crossing of two beams is located in electric field oriented along the laser beam.  The detection of optical rotation (either using simple polarimetry, or phase-sensitive techniques) happens on the output/transmission of the cavity. Let us discuss the ultimate ICAS achievements necessary for the proposed $\mathcal{P}$,$\mathcal{T}$-odd Faraday experiments. In Ref. \cite{Boug14} a possibility to have an optical path length of about 100 km was considered. For a cavity of 1 m length this means $10^5$ passes of the light inside the cavity and $10^5$ reflections of the light from the mirrors. For an atomic/molecular beam-based experiment, typical optical interaction path-lengths are of about 1 cm, i.e. 100 times smaller. However, in another ICAS experiment \cite{Baev99} an optical path length of $7\times 10^4$ km for a cavity of the same size as in Ref. \cite{Boug14} was reported. This means that 700 times enlargement of the light passes number inside a cavity may become realistic. Another important property of ICAS experiments is the sensitivity of rotation angle measurement. Using a cavity-enhanced scheme a shot-noise-limited birefringence-phase-shift sensitivity at the $3\times 10^{-13}$ rad level was demonstrated \cite{Dur10}. We consider the above mentioned parameters used in ICAS experiments to estimate the realizability of the proposed $\mathcal{P}$,$\mathcal{T}$-odd Faraday ICAS experiment for the search of the $\mathcal{P}$,$\mathcal{T}$-odd interactions in atomic (molecular) physics.

\section{Theoretical simulation of ICAS $\mathcal{P}$,$\mathcal{T}$-odd Faraday experiment on atoms and molecules}
The $\mathcal{P}$,$\mathcal{T}$-odd Faraday effect manifests itself as the circular birefringence arising from the light propagating through a medium in an external electric field when the $\mathcal{P}$,$\mathcal{T}$-odd interactions are taken into account. Its origin is the same as for the ordinary Faraday effect in an external magnetic field. In the magnetic field the atomic levels are split into Zeeman sublevels. Then the transitions between two atomic states with emission (absorption) of the right (left) circularly polarized photons correspond to the different frequencies since they occur between different Zeeman sublevels. This causes the birefringence, i.e. different refractive indices $n^{\pm}$ for the right and left photons. The same happens in an external electric field with an account for the $\mathcal{P}$,$\mathcal{T}$-odd interactions. In this case the level splitting magnitude is proportional to the linear Stark shift $S^{\text{EDM}}$. The rotation angle $\psi(\omega)$ of the light polarization plane for any type of birefringence looks like
\begin{equation}
 \label{1}
\psi(\omega)=\pi \frac{l}{\lambda} \text{Re} \left[n^+(\omega)-n^-(\omega)\right]
\end{equation}
where $l$ is the optical path length, $\omega$ is the transition frequency and $\lambda$ is the corresponding wavelength. In the $\mathcal{P}$,$\mathcal{T}$-odd Faraday rotation case \cite{Chub18}
\begin{equation}
 \label{2}
\text{Re} \left[n^+(\omega)-n^-(\omega)\right]= \frac{d}{d\omega} n(\omega) S^{\text{EDM}}.
\end{equation}
The linear Stark shift of atomic levels can be presented as
\begin{equation}
 \label{3}
  S^{\text{EDM}} = R_d d_e \mathcal{E}_{\text{ext}},
\end{equation}
where $R_d$  is the $e$EDM enhancement coefficient in an atom and $\mathcal{E}_{\text{ext}}$ is an external electric field. In case of the  $\mathcal{P}$,$\mathcal{T}$-odd electron-nucleus interaction the value of $d_e$ in \Eq{3} should be replaced by $d_e^{\text{eqv}}$. It is convenient to attribute the enhancement coefficient to $\mathcal{E}_{\text{ext}}$ and to introduce an effective electric field $\mathcal{E}_{\text{eff}}= R_d \mathcal{E}_{\text{ext}}$, which is also commonly used parameter in the molecular case. To extract the $d_e$ value from the experimental data it is necessary to know the value of $\mathcal{E}_{\text{eff}}$ which cannot be measured and should be calculated (see, e.g. \cite{Tit06,Skrip17}). 

In the present paper the effective electric fields in the YbF molecule were calculated within the relativistic Fock-Space coupled cluster method with single and double amplitudes \cite{Vis02} using the Dirac-Coulomb Hamiltonian. To calculate effective electric fields in the PbF molecule we used the relativistic coupled cluster with single, double and noniterative triple cluster amplitudes method. For both molecules all electrons were included in the correlation treatment. For Pb the augmented all-electron triple-zeta AAETZ~\cite{Dyall:06} basis set was used. For Yb and F atoms the all-electron triple-zeta AETZ \cite{Dyall:16,Dyall:10,Dyall:12} basis sets were used. The theoretical uncertainty of these calculations can be estimated as 5\%. The values of $\mathcal{E}_{\text{eff}}$ for the ground electronic states are in good agreement with previous studies \cite{Skripnikov:14c,Sudip:15,Mosyagin:98,Quiney:98,Kozlov:97c,Parpia:98,Abe:14}. 

The rotation signal $R(\omega)$ in the experiment reads
\begin{equation}
\label{6}
R(\omega)= \psi(\omega)T(\omega),
\end{equation}
where $T(\omega)$ is the light transmission. The transmission related to the intracavity losses obeys the Beer-Lambert law
\begin{equation}
\label{7}
T(\omega)=e^{- l/L(\omega)},
\end{equation}
where  $L(\omega)$ is the absorption length. Here we do not consider transmission from a cavity (mirrors) that is used to enhance the interaction optical path length. It changes as a function of intracavity losses and strongly depends on cavity parameters.

Expressed via the spectral characteristics of the resonance absorption line the rotation signal reads
\begin{eqnarray}
 \label{8}
R (\omega) & = & \frac{2\pi^2}{3} \frac{l}{\lambda} \rho e^2  |\langle i |  \bm{r}| f \rangle|^2 \frac{h(u,v)}{\hbar\Gamma_D}  \nonumber
\\  & \times &    \frac{2d_e (\mathcal{E}^i_{\text{eff}}+\mathcal{E}^f_{\text{eff}})}{\Gamma_D}e^{- \rho
 \sigma(\omega)l},
\end{eqnarray}
\begin{equation}
\label{9}
\omega=\omega_0+ \Delta \omega,
\end{equation}
\begin{equation}
\label{10}
\sigma(\omega)=4\pi \frac{\omega_0}{\Gamma_D} f(u,v) \frac{e^2|\langle i |\bm{r} | f\rangle |^2}{3\hbar c}. 
\end{equation}
Here $\rho$ is the atomic (molecular) number density, $| i \rangle$ and $| f \rangle$
are the initial and final states for the resonance transition, $\bm{r}$ is the electron radius-vector, $\Gamma_D$ is the Doppler width, $\mathcal{E}^i_{\text{eff}}$ and $\mathcal{E}^f_{\text{eff}}$ are the effective fields for initial and final states with a positive sign of the total electron angular momentum projection, $\sigma(\omega)$ is the absorption cross-section. $\omega_0$ is the resonance transition frequency, $\Delta \omega$ is the detuning frequency; $\hbar$, $c$ are the reduced Planck constant and the speed of light. For M1 transitions the factor $e^2|\langle i |  \bm{r}| f \rangle|^2$ should be replaced by $\mu_0^2 |\langle i |  \bm{l}-g_S\bm{s}| f\rangle|^2$ where  $\bm{s}$, $\bm{l}$ are the spin and orbital electron angular momenta operators, respectively, $g_S=-2.0023$ is a free-electron $g$ factor and $\mu_0$ is the Bohr magneton.
The functions $g(u,v)$, $f(u,v)$, $h(u,v)$ and the variables $u$, $v$ are connected with the Voigt description of the spectral line profile \cite{Khrip91}
\begin{equation}
\label{11}
 g(u,v) = \text{Im} \; \mathcal{F} (u,v),
\end{equation}
\begin{equation}
\label{12}
f(u,v)= \text{Re}\; \mathcal{F} (u,v) ,
\end{equation}
\begin{equation}
\label{13}
\mathcal{F} (u,v) = \sqrt{\pi} e^{-(u+iv)^2} \left[ 1- \text{Erf} (-i(u+iv)) \right]
\end{equation}
where $\text{Erf}(z)$ is the error function,
\begin{equation}
\label{14}
u=\frac{\Delta\omega}{\Gamma_D}
\end{equation}
and
\begin{equation}
\label{15}
v=\frac{\Gamma_n}{2\Gamma_D}.
\end{equation}
$\Gamma_n$ is the natural width. Finally,
\begin{equation}
\label{16}
h(u,v)=\frac{d}{du} g(u,v).
\end{equation}
 $f(u,v)$ defines the behavior of absorption and has its maximum at $\omega_0$. The maximum of $h(u,v)$ also coincides with $\omega_0$. However, it has the second maximum \cite{Chub18}, what allows to observe the $\mathcal{P}$,$\mathcal{T}$-odd Faraday effect off-resonance, in the region where absorption is very small. For $u \gg 1$ and $v \ll 1$ in asymptotics $h(u,v) \approx 1/u^2$ and $f(u,v)\approx v/u^2$. Then the signal as a function of parameter $\rho l/u^2$ behaves as $R \sim \frac{\rho l}{u^2} e^{-\sigma(\omega_0)v\rho l/u^2}$ and has its maximum at fixed value of $\rho l/u^2 \approx 1/(\sigma(\omega_0)v)$. The further enhancement of the signal $R$ as a function of the column density $\rho l$ and $u$ is impossible. The cooling of atoms and molecules cannot improve the maximum $\mathcal{P}$,$\mathcal{T}$-odd signal value and the maximum rotation angle since $\psi\sim 1/(\Gamma_D)^2 \sim 1/T$ and the absorption coefficient is proportional in asymptotics to $1/(\Gamma_D)^2 \sim 1/T$. Here $T$ is the temperature in Kelvin. In our following estimates we fix $u = 5$. Plotting the signal $R$ (\Eq{8}) as a function of the detuning $u$ with the exact functions $h(u,v)$ and $f(u,v)$ shows that at such detuning (and also larger values of $u$)  $R$ already achieves its maximum. The real observed quantity is the rotation angle but its growth is limited by the absorption. The shot-noise limit for any polarimeter is proportional to the square root of the number of detected photons. Then the signal-to-noise ratio and, correspondingly, statistical sensitivity ($R(\omega)/\sqrt{T(\omega)}$) are optimal when $\frac{d}{dl}\left(l e^{-l /2L(\omega)}\right)=0$, i.e. when $l=2L(\omega)$. It is the case when $T(\omega)$ drops by a factor of $e^2\approx 10$ and the light still can be observed. Here we do not consider the hyperfine structure. If the hyperfine structure is resolved, it does not change the order-of-magnitude estimate for rotation angle. However, the choice of certain hyperfine levels depends on particular experiment. Theoretical simulation of the ICAS $\mathcal{P}$,$\mathcal{T}$-odd Faraday experiment based on above-mentioned circumstances with current $e$EDM limit is presented in Table I. Below we consider some details.

\begin{table} [h!]
\caption{The results for the $\mathcal{P}$,$\mathcal{T}$-odd Faraday optical rotation theoretical simulation for
different molecular and atomic species. The rotation angle $\psi$ corresponds to the present $e$EDM bound $d_e<1.1\times 10^{-29}$ $e$ cm \cite{ACME18}.}
\tabcolsep=0.01cm
\scalebox{0.9}{\begin{tabular}{ccccccc}
\hline\hline  Atom, & Transition & Wavelength  & Linewidth  & Effective field & Column density & Rotation angle   \\
Molecule &  & $\lambda$, nm & $\Gamma_n$, s$^{-1}$ & $\mathcal{E}_{\text{eff}}$, GV/cm & $\rho l$, cm$^{-2}$ & $\psi$, rad \\
\hline 
YbF(YbOH)  &  X $^2\Sigma_{1/2}\rightarrow$ A1 $^2\Pi_{1/2}$ & 552 (577) & $3.6 \times 10^7$  &  
 -23.6(X), 3.1(A1) & 

$3.0\times 10^{12}$ & $5.0 \times 10^{-11}$ \\
PbF & X1 $^2\Pi_{1/2} \rightarrow$ X2 $^2\Pi_{3/2}$ & 1210 & $2.7 \times 10^3$  & 
38.0(X1), 9.3(X2) &
 $1.9\times 10^{19}$ & $1.3 \times 10^{-6}$ \\
Tl & $6p_{1/2}\rightarrow 6p_{3/2}$ & 1283 & 4 & $-5.26\times 10^{-2}$($6p_{1/2}$), $7\times 10^{-4}$($6p_{3/2}$)$^a$ & $1.0\times 10^{22}$ & $2.1 \times 10^{-9}$ \\
Pb & $6p^2 (1/2,1/2)_0 \rightarrow 6p^2(3/2,1/2)_1$ & 1279 & 7  & 0 ($(1/2,1/2)_0$), $2.34 \times 10^{-2}$($(3/2,1/2)_1$) $^a$ &  $1.0\times 10^{22}$ & $6.2 \times 10^{-10}$ \\
ThO & X $^1\Sigma_0\rightarrow$  H $^3\Delta_1$ & 1810 & $5\times 10^{2}$ & 
0(X), 80(H) &
  $9.0\times 10^{19}$ & $1.3 \times 10^{-5}$ \\
\hline \hline
\end{tabular}}
$^a$ $\mathcal{E}_{\text{eff}}$ corresponds to the external electric field equal to 100 kV/cm.
\label{table:1}
\end{table}

1) We begin with the X $^2\Sigma_{1/2}\rightarrow$ A1 $^2\Pi_{1/2}$ transition ($\lambda=552$ nm) in the YbF molecule. The natural linewidth $\Gamma_n=3.6 \times 10^7$ s$^{-1}$ \cite{Alm17}. In Ref. \cite{Alm17} the transverse temperature of the supersonic YbF beam was reported to be about 1 K then the transverse Doppler width $\Gamma_D= 1.1 \times 10^{8}$ s$^{-1}$. Our calculations give the effective electric field values: $\mathcal{E}_{\text{eff}} (^2\Sigma_{1/2,+1/2}) = -23.6$ GV/cm and $\mathcal{E}_{\text{eff}} (^2\Pi_{1/2,+1/2}) = 3.1$ GV/cm. From Fig. 1 one can see that the maximum signal $R_{\text{max}}=3.5\times 10^{-12}$ rad corresponds to $\rho l=6\times 10^{11}$ cm$^{-2}$. The optimal situation when the transmission drops approximately by a factor of 10 corresponds to $\rho l =3.0\times 10^{12}$ cm$^{-2}$. Such column densities can be achieved in a cavity \cite{Boug14} with an atomic (molecular) beam of approximately 1 cm diameter. Then the maximum $\mathcal{P}$,$\mathcal{T}$-odd rotation angle $\psi_{\text{max}} = 5.0 \times 10^{-11}$ rad.
 \begin{figure}[h!]
\begin{center}
\includegraphics[width=7.0 cm]{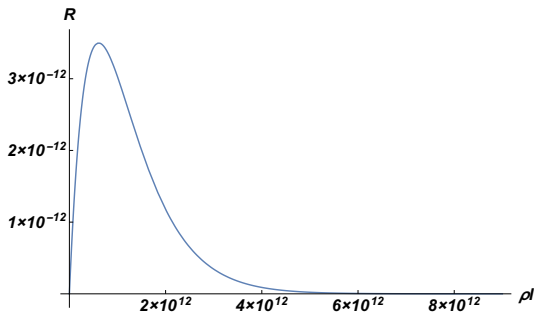}
\end{center}
\caption{\label{f:1} {Dependence of the $\mathcal{P}$,$\mathcal{T}$-odd Faraday signal $R$ (in rad) on column density $\rho l$ (in cm$^{-2}$) for the X $^2\Sigma_{1/2} \rightarrow$ A1 $^2\Pi_{1/2}$ transition in YbF. The present limit $d_e \approx 10^{-29}$ $e$ cm is assumed.}}
\end{figure}
YbOH can be regarded as analogue of YbF for the $e$EDM search ($\lambda=577$ nm). The effective electric fields are approximately the same \cite{Gaul19,Den19,Pras19,Mais19}. To polarize the YbOH molecule one needs small electric field ($\mathcal{E}_{\text{sat}} \approx 30$ V/cm), so it is much easier to deal with it. The YbF molecule can be polarized only by about 50\% at achievable lab fields \cite{Hud11,Hud01}. Taking into account the record sensitivity $3\times 10^{-13}$ rad, YbF and YbOH appear to be good candidates for improving the $e$EDM bound by 2 orders of magnitude.

2) The most promising candidate for the ICAS $\mathcal{P}$,$\mathcal{T}$-odd Faraday experiment with diatomic molecules is the PbF molecule with the X1 $^2\Pi_{1/2} \rightarrow$ X2 $^2\Pi_{3/2}$ transition ($\lambda=1210$ nm). The natural linewidth of the X2 state $\Gamma_n=2.7 \times 10^3$ s$^{-1}$ \cite{Das02}. Similar to the YbF case, for PbF beam we adopt the transverse temperature of 1 K and the transverse $\Gamma_D= 4.5 \times 10^{7}$ s$^{-1}$. Our calculations give the following effective electric fields values: $\mathcal{E}_{\text{eff}} (^2\Pi_{1/2,+1/2}) = 38$ GV/cm and $\mathcal{E}_{\text{eff}} (^2\Pi_{3/2,+1/2}) = 9.3$ GV/cm.  For the optimal signal-to-noise ratio case $\rho l=1.9\times 10^{19}$ cm$^{-2}$, then $\psi_{\text{max}} = 1.3 \times 10^{-6}$ rad. Such column densities cannot be reached in a beam-based experiment using a cavity \cite{Boug14}. In Refs. \cite{Pat07,Hut12} the continuous flux intensity $\Phi$ for molecular (atomic) beams was reported to be $\sim 10^{14} - 10^{15}$ mol (at) s$^{-1}$ sr$^{-1}$.  $\rho=\frac{\Phi}{r^2 v_0}$ where $r$ is the distance from the beam source and $v_0$ is the mean beam speed. For $v_0 \sim 10^{4}$ cm/s and $r\sim 0.1$ cm one obtains $\rho\sim 10^{13}$ cm$^{-3}$.  However, for the optical path length with 700 times enlargement of $l$ \cite{Baev99} the same size of rotation angle can be reached with $\rho=2.7\times 10^{11}$ cm$^{-3}$. As a result, PbF is a good candidate for (optimistically) improving the $e$EDM bound by 6-7 orders of magnitude.

3) The next candidate is the Tl atom with $6p_{1/2} \rightarrow 6p_{3/2}$ transition ($\lambda=1283$ nm). The natural linewidth of the metastable $ 6p_{3/2}$ state $\Gamma_n=4$ s$^{-1}$ \cite{Rad85}. The enhancement coefficients were calculated in Ref.\cite{Chub18}. For Tl beam we adopt the transverse $\Gamma_D= 3.0 \times 10^{7}$ s$^{-1}$ \cite{Fan11}. For the optimal case $\rho l=5.0\times 10^{24}$ cm$^{-2}$, then $\psi_{\text{max}} = 1.1 \times 10^{-6}$ rad. Such a large difference between the results \cite{Chub18} and this result is due to neglect by the collisional broadening for the atomic beam case. Such column densities as $\rho l = 5.0 \times 10^{24}$ cm $^{-2}$ are not achievable in an atomic beam-based experiment. For the path length reported in Ref.\cite{Baev99}  $\rho l = 10^{22}$ cm $^{-2}$ can be achieved. It corresponds to $\psi =2.1 \times 10^{-9}$ rad listed in Table I. However, since at this value the rotation angle does not reach its ``saturated'' maximum $\psi_{\text{max}} =1.1 \times 10^{-6}$ rad, it is possible to increase the angle by reducing of the transverse temperature (the rotation angle is proportional to $1/\Gamma_D \sim 1/T$). In such a way, reducing of the transverse temperature from 1 K to several mK would establish the maximum value $\psi_{\text{max}} = 9.0 \times 10^{-7}$ rad. 

4) Consider now the $6p^2 (1/2,1/2)_0 \rightarrow 6p^2(3/2,1/2)_1$ transition in Pb atom ($\lambda=1279$ nm). The natural linewidth of the metastable $ 6p_{3/2}$ state $\Gamma_n=7$ s$^{-1}$ \cite{Rad85}. The enhancement coefficients were calculated in \cite{Chub18}. For the Pb beam ($T=1$ K) the transverse $\Gamma_D= 4.2 \times 10^{7}$ s$^{-1}$. For the optimal case $\rho l=9.0\times 10^{23}$ cm$^{-2}$, then $\psi_{\text{max}} = 6.0 \times 10^{-8}$ rad. Similar to the Tl atom case, such column densities cannot be reached in an atomic beam-based experiment. In Table I the results for achievable in the present ICAS experiments $\rho l = 10^{22}$ cm $^{-2}$ corresponding to $\psi = 6.2 \times 10^{-10}$ rad  are listed. Reducing the Pb beam transverse temperature to tens of mK with $\rho l = 10^{22}$ cm$^{-2}$ would lead to observable $\psi_{\text{max}} = 6.0 \times 10^{-8}$ rad.

5) Consider the X $^1\Sigma_0\rightarrow$ H $^3\Delta_1$ transition ($\lambda=1810$ nm) in ThO. This transition lies in the infrared region so it does not suit for the present ICAS experiments. Nevertheless, it is interesting theoretically to evaluate the possible rotation angle since the best limitation on the $e$EDM was made on ThO. The natural linewidth of the metastable H state $\Gamma_n=5\times 10^{2}$ s$^{-1}$ \cite{ACME18}. The effective electric field for the H state was calculated in \cite{Skripnikov:13c,Skripnikov:15a,Skripnikov:16b,Fleig:16}. For the ThO beam ($T=1$ K) the transverse $\Gamma_D= 2.9 \times 10^{7}$ s$^{-1}$. For the optimal case $\rho l=9.0\times 10^{19}$ cm$^{-2}$, then $\psi_{\text{max}} = 1.3 \times 10^{-5}$ rad.

\section{Conclusion}

The recent most advanced $e$EDM limit obtained in the experiment with ThO is $d_e<1.1\times 10^{-29}$ $e$ cm. In this experiment an electron spin precession in an external electric field is considered and the effect to be measured is proportional to the time spent by a particular molecule in an electric field. In the present paper we suggest another method for observation of such effects \--- a beam-ICAS $\mathcal{P}$,$\mathcal{T}$-odd Faraday experiment with atoms and molecules. A theoretical simulation of the proposed experiment is based on the recently available ICAS parameters
and do not include corresponding specific technical experimental problems. In this experiment it is not necessary to keep longer any separate atom (molecule) in an electric field since the effect is 
accumulated in the laser beam which meets many atoms (molecules). According to our estimates for the PbF molecule and for Tl, Pb atoms the existing $e$EDM bound can be (optimistically) shifted down by 6-7 orders of magnitude. This implies testing of new particles at energy three order of magnitude larger than in Ref.~\cite{ACME18}, i.e. tens of PeV~\cite{ACME18,Engel2013}.

\begin{acknowledgments}
Preparing the paper and the calculations of the $\mathcal{P}$,$\mathcal{T}$-odd Faraday signals, as well as finding optimal parameters for experiment were supported by the Russian Science Foundation grant 17-12-01035. Rest of the work was supported by the Foundation for the advancement of theoretical physics and mathematics ``BASIS'' grant according to the research projects No.~17-15-577-1 and No.~18-1-3-55-1. The authors would like to thank Dr. Timur A. Isaev, Dr. Alexander N. Petrov, and  Dr. Peter Rakitzis for the helpful discussions.
\end{acknowledgments}


\begin{thebibliography}{99}
\section*{References}
\expandafter\ifx\csname natexlab\endcsname\relax\def\natexlab#1{#1}\fi
\expandafter\ifx\csname bibnamefont\endcsname\relax
  \def\bibnamefont#1{#1}\fi
\expandafter\ifx\csname bibfnamefont\endcsname\relax
  \def\bibfnamefont#1{#1}\fi
\expandafter\ifx\csname citenamefont\endcsname\relax
  \def\citenamefont#1{#1}\fi
\expandafter\ifx\csname url\endcsname\relax
  \def\url#1{\texttt{#1}}\fi
\expandafter\ifx\csname urlprefix\endcsname\relax\def\urlprefix{URL }\fi
\providecommand{\bibinfo}[2]{#2}
\providecommand{\eprint}[2][]{\url{#2}}





\bibitem{Khrip91}
I.B. Khriplovich, Parity Nonconservation in Atomic Phenomena, Gordon and Breach, London, 1991

\bibitem{Gin04}
J.S. Ginges and V.V. Flambaum, Phys. Rep. \textbf{397}, 63 (2004)


\bibitem{Saf18}
M.S. Safronova, D. Budker, D. DeMille, D.F.J. Kimball, A. Derevianko, C.W. Clark, Rev. Mod. Phys. \textbf{90}, 025008 (2018)


\bibitem{ACME18}
V. Andreev et al. (ACME collaboration), Nature \textbf{562}, 355 (2018)


\bibitem{Reg02}
B.C. Regan, E.D. Commins, C.J. Schmidt and D. DeMille, Phys. Rev. Lett. \textbf{88}, 071805 (2002)

\bibitem{Hud11}
J.J. Hudson, D.M. Kara, I.J. Smallman, B.E. Sauer,
M. R. Tarbutt and E. A. Hinds, Nature \textbf{473}, 493 (2011)


\bibitem{Cair17}
W.B. Cairncross, D.N. Gresh, M. Grau, K.C. Cossel, T.S. Roussy, Y. Ni, Y. Zhou, J. Ye and E.A. Cornell, Phys.Rev.Lett. \textbf{119}, 153001 (2017)


\bibitem{Liu92}
Z.W. Liu and Hugh P. Kelly, Phys. Rev. A \textbf{45}, R4210(R) (1992)

\bibitem{Dzuba09}
V.A. Dzuba and V.V. Flambaum, Phys.Rev. A \textbf{80}, 062509 (2009)

\bibitem{Por12}
S.G. Porsev, M.S. Safronova and M.G. Kozlov, Phys.Rev.Lett. \textbf{108}, 173001 (2012)

\bibitem{Chub18}
D.V. Chubukov, L.V. Skripnikov and L.N. Labzowsky, Phys. Rev. A \textbf{97}, 062512 (2018)

\bibitem{Quiney:98}
H.M. Quiney, H. Skaane, and I.P. Grant, J.Phys. B: At.Mol.Opt.Phys.  \textbf{31}, 85 (1998)
 
\bibitem{Parpia:98}
F. Parpia, J.Phys. B: At.Mol.Opt.Phys. \textbf{31}, 1409 (1998)

\bibitem{Mosyagin:98}
N.S. Mosyagin, M.G. Kozlov, and A.V. Titov, J.Phys. B: At.Mol.Opt.Phys. \textbf{31}, L763 (1998)


\bibitem{Abe:14}
M. Abe, G. Gopakumar, M. Hada, B.P. Das, H. Tatewaki, and D. Mukherjee, Phys. Rev. A \textbf{90}, 022501 – (2014)



\bibitem{Skripnikov:13c}
L.V. Skripnikov, A.N. Petrov, and A.V. Titov, J.Chem.Phys. \textbf{139}, 221103 (2013)
   
\bibitem{Skripnikov:15a}
L.V. Skripnikov, and A.V. Titov, J.Chem.Phys. \textbf{142}, 024301 (2015)
  
 \bibitem{Skripnikov:16b}
L.V. Skripnikov, J.Chem.Phys. \textbf{145}, 214301 (2016)

\bibitem{Fleig:16}
M. Denis, and T. Fleig, J.Chem.Phys. \textbf{145}, 214307 (2016)

\bibitem{Petrov:07a}
A.N. Petrov, N.S. Mosyagin, T.A. Isaev, and A.V. Titov, Phys.Rev. A \textbf{76}, 030501(R) (2007)

\bibitem{Skripnikov:17c}
L.V. Skripnikov, J.Chem.Phys. \textbf{147}, 021101 (2017)
 
\bibitem{Fleig:17}
 T. Fleig, Phys.Rev. A \textbf{96}, 040502 (2017)
 
\bibitem{Petrov:18b}
A.N. Petrov, L.V. Skripnikov, A.V. Titov, and V.V. Flambaum, Phys.Rev. A \textbf{98}, 042502 (2018)



\bibitem{San75}
P.G.H. Sandars, At.Phys. \textbf{4}, 71 (1975)

\bibitem{Gor79}
V.G. Gorshkov, L.N. Labzowsky and A.N. Moskalev, ZhETF \textbf{76}, 414 (1979)
[Sov. Phys. JETP \textbf{49}, 209 (1979)]

\bibitem{Koz95}
M.G. Kozlov, and L.N. Labzowsky, J.Phys. B: At.Mol.Opt.Phys. \textbf{28}, 1933 (1995)

\bibitem{Bon15}
A.A. Bondarevskaya, D.V. Chubukov, O.Yu. Andreev, E.A. Mistonova, L.N. Labzowsky, G. Plunien, D. Liesen, F. Bosch, J. Phys. B \textbf{48}, 144007 (2015)


\bibitem{Engel2013}
J. Engel, M. J. Ramsey-Musolf, and U. van Kolck, Prog. Part. Nucl. Phys. \textbf{71}, 21 (2013)


\bibitem{Pos14}
M. Pospelov and A. Ritz, Phys. Rev. D \textbf{89}, 056006 (2014)

\bibitem{Chub16}
D.V. Chubukov and L.N. Labzowsky, Phys. Rev. A \textbf{93}, 062503 (2016)

\bibitem{Baran78}
N.B. Baranova, Yu.V. Bogdanov and B.Ya. Zel'dovich, Usp.Fiz.Nauk \textbf{123}, 349 (1977) [Sov.Phys.Usp. \textbf{20}, 1977, (1978)]

\bibitem{Sush78}
O.P. Sushkov,and V.V. Flambaum, ZhETF \textbf{75}, 1208 (1978) [Sov. Phys. JETP \textbf{48}, 608 (1978)]

\bibitem{Bud02}
D. Budker, W. Gawlik, D. Kimball, S.M. Rochester, V. Yashchuk and A. Weis, Rev.Mod.Phys. \textbf{74}, 1153 (2002)

\bibitem{Boug14}
L. Bougas, G.E. Katsoprinakis, W. von Klitzing and T.P. Rakitzis, Phys. Rev. A \textbf{89}, 052127 (2014)

\bibitem{Baev99}
V.M. Baev, T. Latz and P.E. Toschek, Appl. Phys. B \textbf{69}, 171 (1999) 

\bibitem{Dur10}
M. Durand, J. Morville and D. Romanini, Phys. Rev. A \textbf{82}, 031803 (2010)


\bibitem{Chub17}
D.V. Chubukov and L.N. Labzowsky, Phys. Rev. A \textbf{96}, 052105 (2017)

\bibitem{Chub19:1}
D.V. Chubukov, L.V. Skripnikov, L.N. Labzowsky, V.N. Kutuzov, and S.D. Chekhovskoi, Phys.Rev. A \textbf{99}, 052515 (2019)

\bibitem{Chub19:2}
D.V. Chubukov, L.V. Skripnikov, V.N. Kutuzov, S.D. Chekhovskoi, and L.N. Labzowsky, Atoms, 7, 56; doi:10.3390/atoms7020056 (2019)



\bibitem{Lab77}
L.N. Labzovsky, Zh. Eksp. Teor. Fiz. \textbf{73}, 1623 (1977)
[Sov. Phys. JETP \textbf{46}, 853 (1977)]



\bibitem{Koz17}
I. Kozyrev, and N.R. Hutzler, Phys.Rev.Lett. \textbf{119}, 133002 (2017)

\bibitem{Tit06}
A.V. Titov, N.S. Mosyagin, A.N. Petrov, T.A. Isaev and D. DeMille, ``Study of parity violation effects in polar heavy-atom molecule'', In: Recent Advances in the Theory of Chemical and Physical Systems ["Progress in Theoretical Chemistry and Physics" series] v.15, chapt.2, pp. 253-284 (2006)


\bibitem{Skrip17}
L.V. Skripnikov, J.Chem.Phys., \textbf{147}, 021101 (2017)

\bibitem{Vis02}
L. Visscher, E. Eliav and U. Kaldor, J.Chem.Phys. \textbf{115}, 9720 (2002)

\bibitem{Dyall:06}
K.G. Dyall, Theor. Chem. Acc.,  \textbf{115}, 441 (2006)


\bibitem{Dyall:16}
K.G. Dyall, Theor. Chem. Acc.,  \textbf{135}, 128 (2016)


\bibitem{Dyall:10}
A.S.P. Gomes, K.G. Dyall and L. Visscher, Theor. Chem. Acc. \textbf{127}, 369 (2010)

\bibitem{Dyall:12}
K.G. Dyall, Theor. Chem. Acc. \textbf{131}, 1217 (2012)


\bibitem{Skripnikov:14c}
L.V. Skripnikov, A.D. Kudashov, A.N. Petrov, A.V. Titov, Phys.Rev. A \textbf{90}, 064501 (2014)


\bibitem{Sudip:15}
S. Sasmal, H. Pathak, M.K. Nayak, N. Vaval, S. Pal J. Chem. Phys. \textbf{143}, 084119 (2015)


\bibitem{Kozlov:97c}
M.G. Kozlov, J.Phys. B \textbf{30}, L607 (1997)






\bibitem{Alm17}
J.R. Almond, PhD Thesis ``Laser cooling of YbF molecules for an improved measurement of the electron electric dipole moment'', Imperial College London (2017)


\bibitem{Gaul19}
K. Gaul, R. Berger, arXiv:1811.05749 [physics.chem-ph]

\bibitem{Den19}
M. Denis, P.A.B. Haase, R.G.E. Timmermans, E. Eliav, N.R. Hutzler, A. Borschevsky, 	arXiv:1901.02265 [physics.atom-ph]

\bibitem{Pras19}
V.S. Prasannaa, N. Shitara, A. Sakurai, M. Abe, B.P. Das, arXiv:1902.09975 [physics.atom-ph]

\bibitem{Mais19}
D.E. Maison, L.V. Skripnikov, V.V. Flambaum, arXiv:1906.11487 [physics.atom-ph]


\bibitem{Hud01}
J.J. Hudson, PhD Thesis ``Measuring the electric dipole moment of the electron with YbF molecules'', University of Sussex (2001) 

\bibitem{Das02}
K.K. Das, I.D. Petsalakis, H.-P. Liebermann, A.B. Alekseyev, R.J. Buenker, J.Chem.Phys. \textbf{116}, 608 (2002)

\bibitem{Pat07}
D. Patterson, J.M. Doyle, J.Chem.Phys. \textbf{126}, 154307 (2007)

\bibitem{Hut12}
N.R. Hutzler, H.-I Lu, J.M. Doyle, Chem.Rev. \textbf{112}, 4803 (2012)

\bibitem{Rad85}
A.A. Radzig, B.M. Smirnov, ``Reference Data on Atoms, Molecules, and Ions'', Springer (1985)

\bibitem{Fan11}
I. Fan, T.-L. Chen, Y.-S. Liu, Y.-H. Lien, J.-T. Shy, and Y.-W. Liu, Phys.Rev. A \textbf{84}, 042504 (2011)




\end{thebibliography}
\end{document}